# An underground fiber cable discrimination method based on laser interferometer

Dongqi Song[1,2], Guan Wang[1], Zhongwang Pang[1], Bo Wang[1,2]

(1. State Key Laboratory of Precision Space-time Information Sensing Technology, Department of Precision Instrument, Tsinghua University, Beijing 100084, China; 2. Weiyang College, Tsinghua University, Beijing 100084, China)

**Abstract:** The maintenance and repair of optical fiber networks often requires the discrimination of underground cables among a group of buried ones. Presently, the methods commonly used are either inefficient or harmful to the fiber cable. In this paper, we evaluate the effectiveness of fiber optic vibration sensing method on underground fiber cable discrimination. We find that the typical vibration sensing method-distributed acoustic sensing (DAS) is not suitable for fiber cable discrimination. Especially for long distance scenario, due to "frequency grafting" effect, the DAS method will give a false response on the spectrum of knock events during cable discrimination process. In response, we propose an underground fiber cable discrimination method based on laser interferometer, and demonstrate it on the 40-km urban fiber cable connecting Tsinghua University and Yongding Road. It can give an obvious and real response to the knock event, and can be used in practical applications.

**Introduction:**

Optical fiber networks have become one of the largest infrastructures for human utilization, and its daily maintenance is an important guarantee for the normal operation of the network [1-3]. Facing fierce competition in the optical cable operation and maintenance industry, the introduction of new technologies will improve the efficiency and competitiveness of enterprises. Here, we discuss a typical maintenance scenario: fiber cable branching in telecommunication manhole. Considering that the fiber cables in the manhole are laid in different years and operated by different companies, labels on fiber optic cables have become blurred or have fallen off over time, as shown in Fig. 1(a). Construction crews need to determine and cut a certain cable of their own company during branching process. Cutting a wrong cable may cause problems such as extra workload for repairing and even being claimed for compensation.

Consequently, a simple and efficient method for fiber cable discrimination is widely needed in communication industry. A feasible way is to manually apply a vibration on a certain cable and detect it remotely, for example, knocking the cable with common tools that construction crews often carry. We are surprised to learn that the distributed acoustic sensing (DAS) method [4-9], which is widely used in fields such as perimeter security, oil and gas pipeline monitoring, and traffic monitoring [10-20], has not been used to solve this problem. In practice, two methods are mainly in use. One is to pull the cable and feel the force in the nearby communication manhole. It is inefficient and even unpractical for long cables with several manholes along them. The other is to bend the cable and observe the transmission loss using an optical time domain reflectometry (OTDR), but the bending operation may cause irreversible damage to the cable, and is hard to realize on bending resistant optical cables.

In this paper, we analyze the response feature of DAS system on knock events and find that, due to 'frequency grafting' effect, DAS cannot give the correct response. The effect is significant on long-distance tasks. For example, on the 40-km cable connecting Tsinghua and Yongding Road, it is difficult to distinguish the knock events from traffic noise. In response, we demonstrate that fiber optic interferometer can perfectly solve the problem, clearly detect the knocking induced vibration from traffic noise, and can be used as fiber cable discriminator in practical applications.

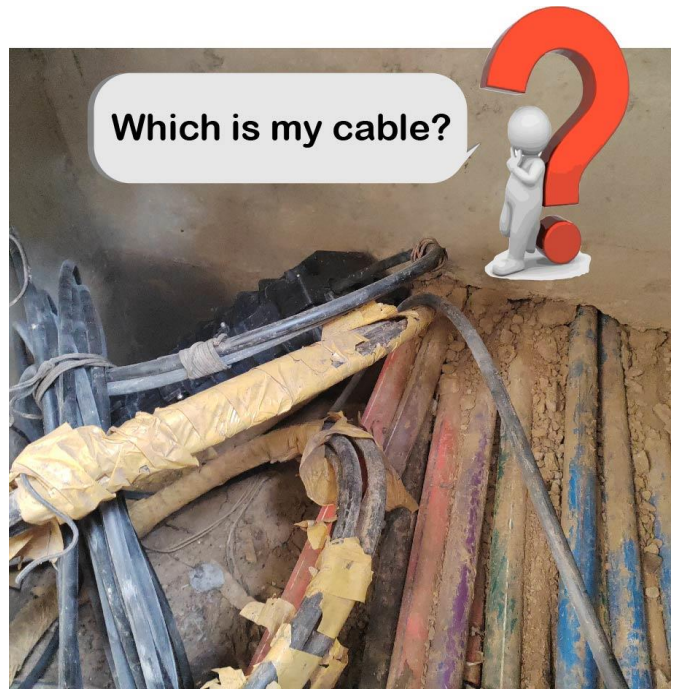

(a)

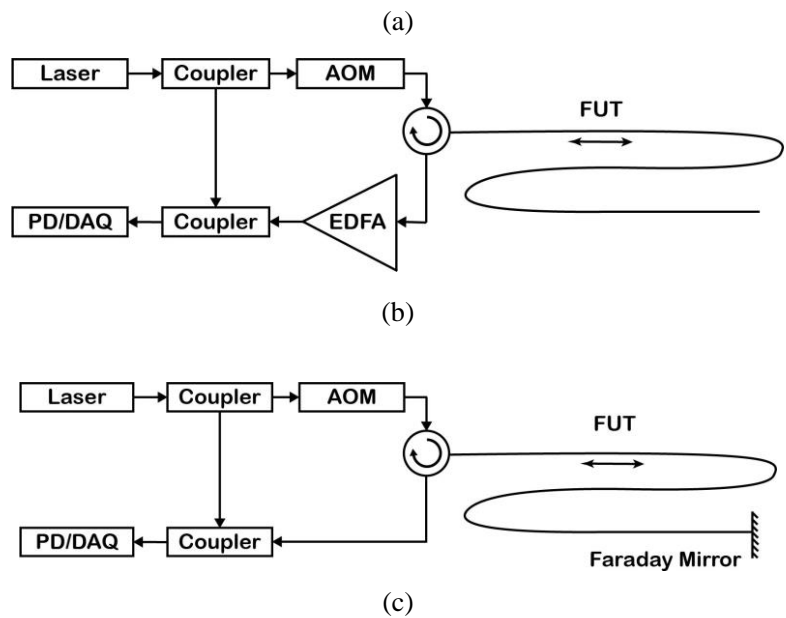

(b)

(c)

**Fig. 1.** (a) Fiber cables under the communication manhole. Crews can hardly find labels on the cable. (b) Schematic diagram of DAS detection system, (c) Schematic diagram of laser interferometer detection system. FUT: Fiber under test.

**Main:**

DAS is a powerful fiber vibration sensing method. It is based on the Rayleigh back-scattering light of the input laser pulse in the cable. As is shown in Fig. 1(b), using an acousto-optic modulator (AOM), the input laser is turned into a series of pulses with a certain repetition rate $f_{rep}$. The pulses are sent into the cable, and the Rayleigh back-scattered pulses from different positions of the cable return at different times. A circulator is used to separate the back-scattered light. After interfering with the input light and phase demodulation, the phase changes caused by vibrations at different times and positions can be detected, realizing distributed sensing. The sampling rate of DAS is equal to the repetition rate of sampling pulse $f_{rep}$, each pulse will conduct one sampling at each position. There is a trade-off between repetition rate $f_{rep}$ and detection distance L of the fiber cable. Only after the previous pulse travels round trips along the whole fiber, can the latter pulse emit to sense vibrations in the next period. Otherwise, there will be overlaps between periods and one cannot relate time delay information with distance [21-22]. Thus, the repetition rate $f_{rep}$ should satisfy

$$f_{rep} \leq \frac{c}{2nL} \tag{1}$$

where n is the refractivity of the fiber, and c is the vacuum light speed. Besides, the vibration information is loaded on the phase of light signal, so the phase change speed of the light signal relates to not only the frequency of the vibration, but also the range of phase change $\Delta\phi$ caused by the vibration. Phase change coefficient $\frac{\Delta\phi}{2\pi}$ (seen as a dimensionless quantity) is used to replace $\Delta\phi$ for convenience. Considering a single-frequency vibration with frequency $f_{vib}$ and phase change coefficient $\frac{\Delta\phi}{2\pi}$, the phase variation caused by the vibration can be written as S(t) = $\frac{\Delta\phi}{2\pi}\sin(2\pi f_{vib}t)$. According to the sampling theorem, the phase difference between two sampling points should be smaller than $\pi$, thus

$$max\left\{\frac{dS(t)}{dt}\right\} * \frac{1}{f_{rep}} = \frac{\Delta\phi}{2\pi} * 2\pi \frac{f_{vib}}{f_{rep}} \leq \pi \tag{2}$$

The relation among repetition rate $f_{rep}$, vibration frequency $f_{vib}$ and phase change coefficient $\Delta\phi/2\pi$ is

$$f_{rep} \geq 2f_{vib}\frac{\Delta\phi}{2\pi} \tag{3}$$

which adds a modification term into the sampling theorem. When the vibration frequency $f_{vib}$ or phase change coefficient $\frac{\Delta\phi}{2\pi}$ is too high, i.e., $2f_{vib}\frac{\Delta\phi}{2\pi} > \frac{c}{2nL}$, repetition rate $f_{rep}$ derived by Eq. (1) can never satisfy the modified sampling theorem Eq. (3), and DAS system cannot give the result correctly.

In phase demodulation process, the inverse trigonometric function is used to get the phase information. Inverse trigonometric function treats two phases the same when they differ by $2\pi$, so it will introduce an uncertainty of $2m\pi$ to the phase result (where m is an integer number and increases with the increase of $f_{vib}\frac{\Delta\phi}{2\pi}$). Normally, when the modified sampling theorem ($f_{rep} \geq 2f_{vib}\frac{\Delta\phi}{2\pi}$) is satisfied, there will be at least two sampling points within each period, and phase change between adjacent sampling points will never be more than $\pi$. Given this assumption, the uncertainty can be avoided, and one can obtain the correct phase. However, when $2f_{vib}\frac{\Delta\phi}{2\pi} > \frac{c}{2nL}$, modified sampling theorem cannot be satisfied. The phase change between two adjacent points can be larger than $\pi$, and the phase result will carry a $2m\pi$ error. In time domain, it seriously resists the detected phase change to a very small range. In frequency domain, it leads to harmonics as high frequency noises, and some harmonics higher than repetition rate $f_{rep}$ will be undersampled as low frequency noises. The phenomenon of moving and noise in frequency domain is like grafting a frequency component to many other places, and we call it 'frequency grafting'.

Previous studies show that the vibration frequency of manual knocking is mainly between 800 Hz and 2000 Hz [23-26]. Considering that the force of knocking is hard to control, the phase change coefficient $\frac{\Delta\phi}{2\pi}$ may be large, and the modified sampling theorem ($2f_{vib}\frac{\Delta\phi}{2pi} \leq \frac{c}{2nL}$) may still be hard to satisfy. So, we first conduct experiments to see what will happen with 'frequency grafting'. Experiments are conducted on a 1-km cable in lab, and a piezoelectric ceramic is used to generate a single-frequency vibration signal with $f_{vib} = 1$ kHz. Repetition rate is set as $f_{rep} = 62.5$ kHz, so when there are harmonics of $f_{vib} = 1$ kHz, the peak at 500 Hz will appear due to undersampling (generated by the 63rd harmonic as 63×1 kHz-62.5 kHz = 500 Hz). The phase change coefficient $\frac{\Delta\phi}{2\pi}$ gradually increases from 3 to 300 and the results are shown as black lines in Fig. 2. Simulations are also conducted under the same conditions, and red lines are the results in time domain. When $\frac{\Delta\phi}{2\pi} = 3$, the single frequency signal can be solved perfectly as shown in Fig. 2(a). For $\frac{\Delta\phi}{2\pi} = 10$, the experiment result also agrees with simulation, with only several sharp peaks missing. Errors of $2m\pi$ appear in time domain and resist the detected phase change range to be even smaller than Fig. 2(a) with $\frac{\Delta\phi}{2\pi} = 3$. Observing the frequency spectrum, the peak at 500 Hz appears as analyzed above, indicating that 'frequency grafting' is happening. When $\frac{\Delta\phi}{2\pi}$ is more than 30, the results are similar, and result with $\frac{\Delta\phi}{2\pi} = 300$ is chosen as an example, as shown in Fig. 2(c). Since $\frac{\Delta\phi}{2\pi}$ is too large, the experiment result is seriously distorted. The shape of wave peaks and valleys remains the same as that in simulation, but the periodicity is almost destroyed by $2m\pi$ errors. The loss of periodicity reflects in frequency domain that, the 1 kHz characteristic frequency peak is missing, and several peaks of lower and higher frequency components appear. The results above show that 'frequency grafting' will weaken and even kill the component of main frequency $f_{vib}$, making the signal more likely to be submerged by traffic noise [27-31] in real applications.

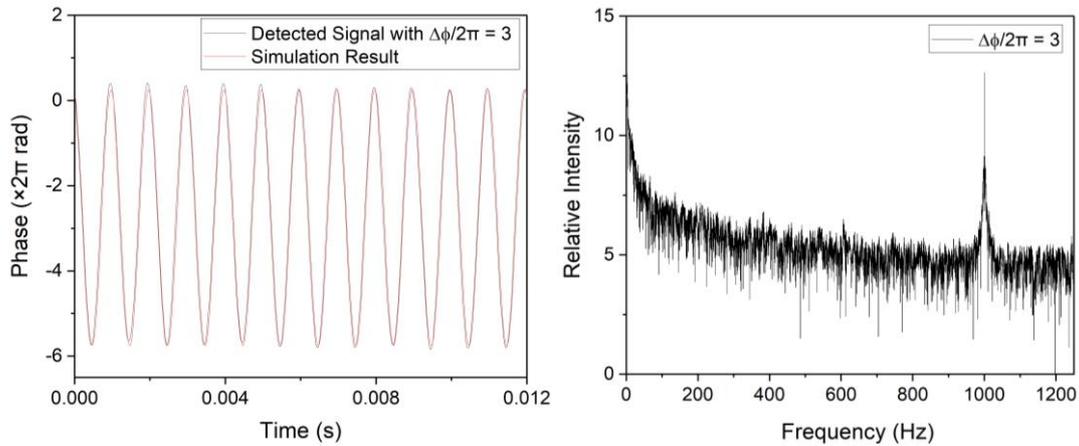

(a)

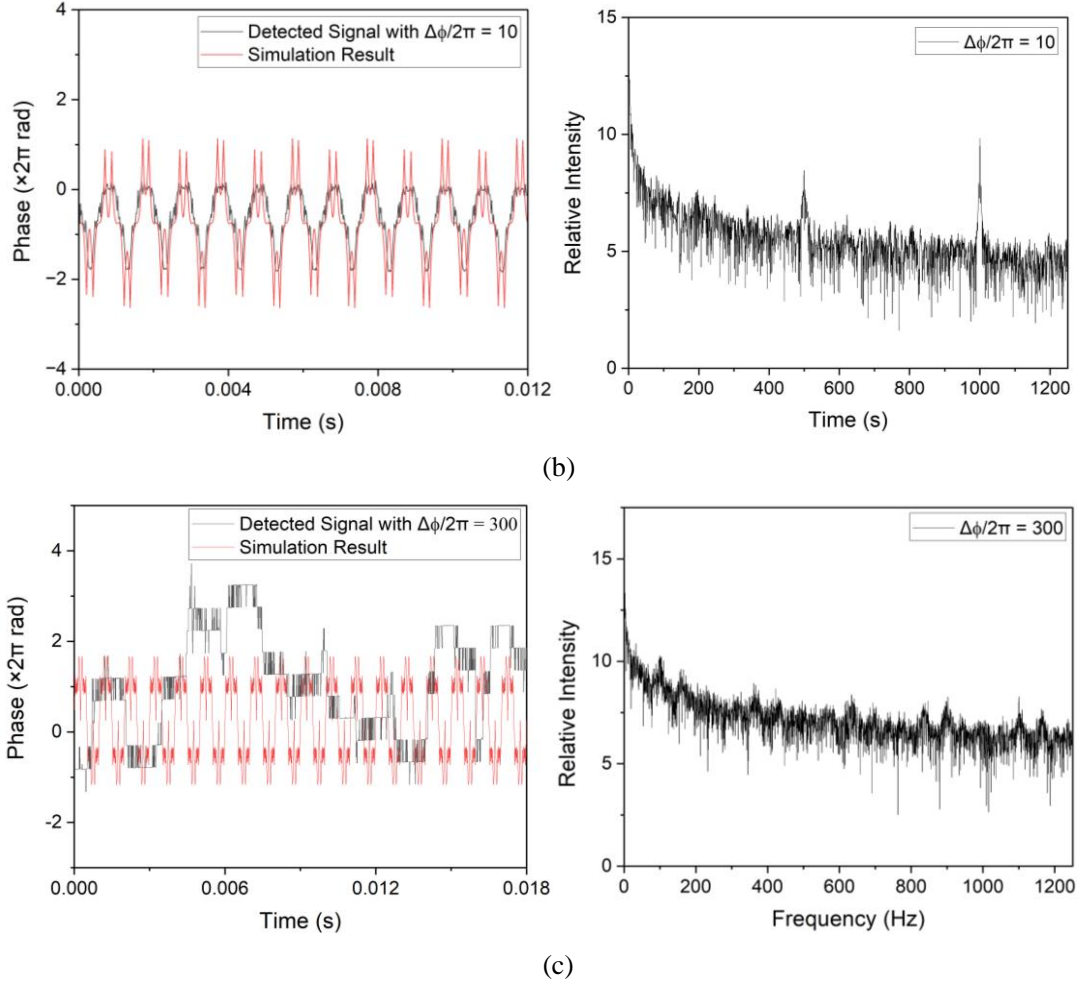

**Fig. 2.** The single-frequency vibration detection results. (a), (b), and (c) represent different phase change coefficient $\frac{\Delta\phi}{2\pi}$ of 3, 10, and 300, respectively. Black lines are experiment results, while red ones are simulation results. 'Frequency grafting' is taking effect, with 1 kHz component weakening and noises of other frequencies gradually appearing.

As discussed above, the restriction of DAS ability is due to the discontinuity of input, which seriously restricts the repetition rate $f_{rep}$, especially when detected distance L reaches tens of kilometers. There are two mainstream methods of distributed fiber optical sensing. One is DAS system, and the other is based on laser interferometer. [32-36] The main difference between DAS and laser interferometer is that, laser interferometer uses continuous light input. As shown in Fig. 1(c), the frequency of laser is shifted by a certain frequency using AOM. Then it is sent into the cable. On the farther end of cable, a Faraday mirror is used to reflect the light back. A circulator is used to separate the round-trip light from the input end of cable, and then interfere it with the source light. After demodulation, one can obtain the phase information. The main advantage of laser interferometer is that, the input light is continuous so there is no trade-off between distance and sampling rate. Higher sampling rate $f_{sam}$ makes it possible to meet the modulated sampling theorem ($f_{sam} \geq 2 * f_{vib} * \frac{\Delta\phi}{2\pi}$) and thus can get the vibration information without distortion.

Based on the analysis, we propose an underground fiber cable discrimination method using laser interferometer. As a demonstration, both DAS system and laser interferometer system are tested on the 1-km cable in lab and a hammer is used to generate a manual signal for testing. The repetition frequency of DAS system $f_{rep}$ is set as 80 kHz. The driving frequency of AOM in laser interferometer is set as 80 MHz, and sampling rate as 1.2 MHz. The detection results of both systems are shown in Fig. 3. The knock signal can be observed clearly in the result of laser interferometer with phase change coefficient $\frac{\Delta\phi}{2\pi}$ around 300. In frequency spectrum of laser interferometer, the peak appears at $f_{vib} \approx 1$ kHz. Under this condition, the modulated sampling theorem for DAS is not satisfied, and the peak around 1 kHz will be 'grafted'. In time domain, the phase vibration range of DAS is seriously resisted to around $0.6\times 2\pi$ as shown in Fig. 3(c). In frequency domain, the 1 kHz characteristic frequency peak is seriously weakened, while peaks of lower frequencies appear as noises, as shown in Fig. 3(d). Similar arising of lower frequency peaks can be found in the $\frac{\Delta\phi}{2\pi} = 300$ experiment as shown in Fig. 2(c). In summary, the test results on 1-km cable verify that laser interferometer can handle the task, while the result of DAS will be influenced by 'frequency grafting' effect and will be distorted, especially on long-distance cable.

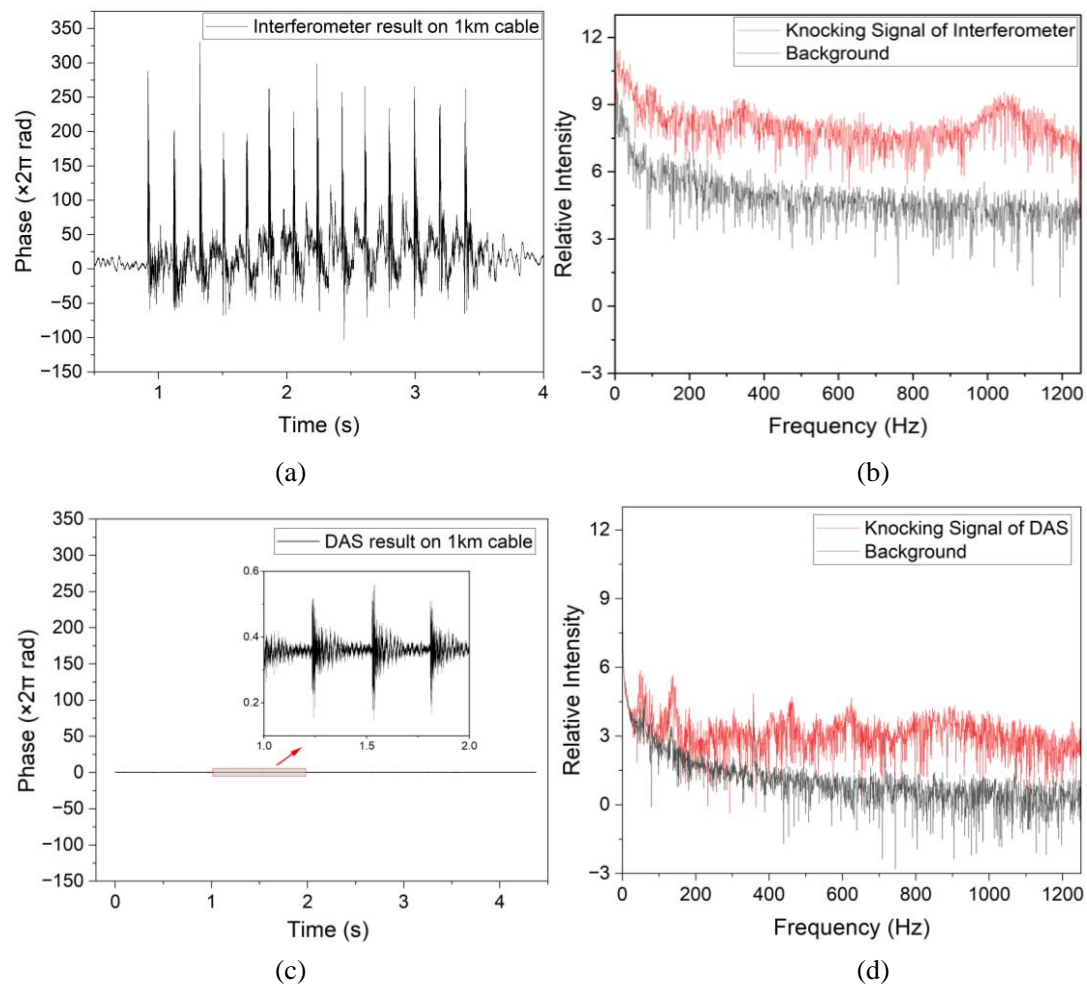

**Fig. 3.** Results of laser interferometer on 1-km cable in lab in time (a) and frequency (b) domain. Vibration signal causes phase change with range of around $300 * 2\pi$ and the characteristic frequency of 800-1200 Hz can be seen clearly. Results of DAS on 1-km cable in lab in time (c) and frequency (d) domain. The amplitude of signal is seriously weakened and the peak of characteristic frequency also decreases to a very low level, while several peaks of lower frequencies appear as noises.

We then take a step forward and test the laser interferometer system on 40-km cable from Tsinghua University to Yongding Road. A hammer is taken to knock the fiber, and the round-trip light is observed to get the vibration information. The route map is shown in Fig. 4(a). The driving frequency of AOM and sampling rate are the same as above. The result is shown in Fig. 4(b) and Fig. 4(c). There are 12 peaks in time domain, indicating 12 times of knockings on the cable, and the phase change range caused by knocking is around $300\times2\pi$. The peak in frequency spectrum is around 1 kHz. The result verifies that laser interferometer system can handle the discrimination task with distance up to 40 km. DAS system is also tested on the 40-km cable and the knockings cannot be distinguished from traffic noise.

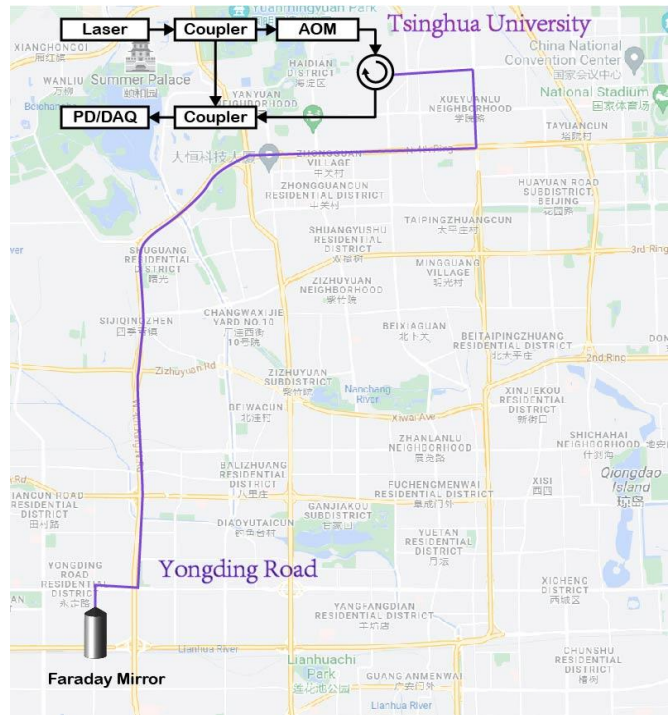

(a)

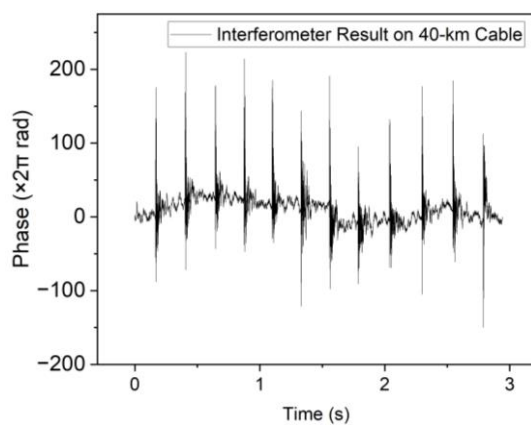

(b)

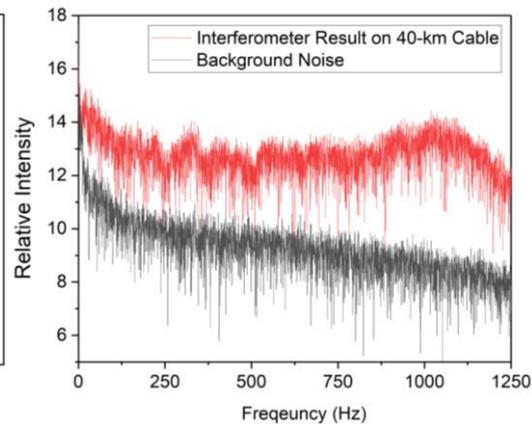

(c)

**Fig. 4.** (a) Route map of 40-km cable under test from Tsinghua University to Yongding Road. (b) Result of laser interferometer on the 40-km cable. There are 12 peaks indicating 12 knockings with a hammer. The Range of phase change detected is around $300\times2\pi$. (c) Result of laser interferometer in frequency domain. Red line describes the background noise and black line the spectrum for knocking. The peak in frequency spectrum appears

in the 800-1200 Hz band.

**Conclusion:**

In this paper, we find that the typical vibration method-DAS fails in the long distance cable discrimination scenario. Affected by "frequency moving" effect, the DAS method will give a false response on the spectrum to the knock event during cable discrimination process. In response, we propose an underground fiber cable discrimination method based on laser interferometer, and demonstrate it on the 40-km urban fiber cable connecting Tsinghua University and Yongding Road. It can give a clear and real response to the knock event, and is promising to be used in practical applications.

**Funding acknowledgement:**
National Natural Science Foundation of China (61971259); National Natural Science Foundation of China (62171249); National Key Research and Development Program of China (2021YFA1402102); Tsinghua Initiative Scientific Research Program.